\documentclass[conference,onecolumn,12pt]{IEEEtran}
\usepackage{cite}
\usepackage{graphicx}

\begin{document}
\title{Fast Simulations of Gravitational Many-body Problem on RV770 GPU}
\author{
\authorblockN{Kazuki Fujiwara and Naohito Nakasato}
\authorblockA{Department of Computer Science and Engineering\\
University of Aizu\\
Aizu-Wakamatsu, Fukushima 965-0815, Japan\\
Contact Email: nakasato@u-aizu.ac.jp}
}

\maketitle

\begin{abstract}
The gravitational many-body problem is a problem concerning 
the movement of bodies, which are interacting through gravity.
However, solving the gravitational many-body problem with a CPU takes a lot of time
due to $O(N^2)$ computational complexity.
In this paper, we show how to speed-up the gravitational many-body problem by using GPU.
After extensive optimizations,  the peak performance obtained so far
is $\sim 1$ Tflops. 
\end{abstract}

\IEEEpeerreviewmaketitle

\section{Introduction}
A gravitational many-body simulation technique
is fundamental in astrophysical simulations
because gravity force drives the structure formation in the universe.
Length scales arisen in the structure formation
range from less than 1 cm at aggregation of dust to more than $10^{24}$ cm 
at formation of cosmological structure.
In all scales, gravity is a key physical process to understand the structure formation.
The reason behind this is long-range nature of gravity.

Suppose we simulate the structure formation with $N$ particles, 
a flow of a many-body simulation is as follows.
First, we calculate mutual gravity force between $N$ particles then integrate
orbits for $N$ particles and repeat this process as necessary.
Although it is simple, the force-calculation is a challenging task in regarding computational science.
A simple and exact method to do the force-calculation requires $O(N^2)$ computational complexity, 
which is prohibitively compute intensive with large $N$.
The exact force-calculation is necessary in some types of simulations such as
a few-body problems, numerical integration of planets orbiting around a star (e.g., the Solar system), 
and evolution of dense star clusters.
For simulations that do not require exact force, a several approximation techniques
have been proposed \cite{Hockney_1981, Barnes_1986, Greengard_1987}.
The particle-mesh/particle-particle-mesh method \cite{Hockney_1981} and the oct-tree method \cite{Barnes_1986}
reduce the computational complexity of the force-calculation to $O(N {\rm log} N)$.
The fast-multipole method \cite{Greengard_1987} further reduces it to $O(N)$.

An computational technique to evaluate the exact force-calculation rapidly 
is to ask for a help of a special hardware like GRAPE\cite{Sugimoto_1990, Makino_1998}.
Precisely, the exact force-calculation is expressed as following equations;
\begin{eqnarray}
\mbox{\boldmath{$a$}}_{i} &=& \sum_{j=1,j \ne i}^{N} \mbox{\boldmath{$f$}}(\mbox{\boldmath{$x$}}_i, \mbox{\boldmath{$x$}}_j, m_j) = \sum_{j=1,j \ne i}^{N} \frac{m_j (\mbox{\boldmath{$x$}}_i - \mbox{\boldmath{$x$}}_j)}
                {(|\mbox{\boldmath{$x$}}_i - \mbox{\boldmath{$x$}}_j|^2 + \epsilon^2)^{3/2}}, \nonumber \\
p_i &=& \sum_{j=1,j \ne i}^{N} p(\mbox{\boldmath{$x$}}_i, \mbox{\boldmath{$x$}}_j, m_j) =  
         \sum_{j=1,j \ne i}^{N} \frac{m_j}
                {(|\mbox{\boldmath{$x$}}_i - \mbox{\boldmath{$x$}}_j|^2 + \epsilon^2)^{1/2}}, \nonumber
\label{gravity}
\end{eqnarray} 
where $\mbox{\boldmath{$a$}}_{i}$ and $p_i$ are force vector and potential
for a particle $i$, and $\mbox{\boldmath{$x$}}_i$, $m_i$, $\epsilon$ are 
position of a particle, the mass, and a parameter that prevents division by zero, respectively.
It is apparent that force-calculation for each particles are independent.
Therefore, the exact force-calculation is difficult but a massively parallel problem.
In the GRAPE system, they have taken full advantage of this fact by
computing different force in parallel with many computing pipelines.
It is natural to take the same approach to utilize a recent graphic processing unit (GPU), 
which has many number of arithmetic units $\sim ~ 200 - 800$, 
for the exact force-calculation.
The rise of the GPU forces us to re-think a way of parallel computing on it
since a performance of recent GPUs is impressive at $> 1$ Tflops.
Acceleration techniques for the exact force-calculation with GPU
have been already reported (\cite{Nyland_2007} and many others).

In this paper, we report our technique to speed-up the exact force-calculation
on RV770 GPU from AMD/ATi.
As far as we know, our implementation on RV770 GPU running at 750 MHz shows fastest
performance of $\sim 1$ Tflops thanks to efficient cache architecture of RV770 GPU. 
Furthermore, a loop-unrolling technique is highly effective RV770 GPU.
In the following sections, we briefly describe our method, implementation and
performance.

\section{Our Computing System with RV770 GPU}
Our computing system used in the present paper consists of a host computer and 
an extension board.
A main component of the extension board is a GPU processor that 
acts as an accelerator attached to the host computer.

\subsection{Architecture of RV770 GPU}
RV770 processor from AMD/ATi is the company's latest GPU (R700 architecture)
with many enhancements for general purpose computing on GPU (GPGPU).
It has 800 arithmetic units (called a stream core), each of which is capable of executing
single precision floating-point (FP) multiply-add in one cycle.
At the time of writing, the fastest RV770 processor is running at 750 MHz
and offers a peak performance of $800 \times 2 \times 750 \times 10^6 = 1.2$ Tflops.
Internally, there are two types of the stream cores in the processor.
One is a simple stream core that can execute only a FP multiply-add
and integer operations and operates on 32 bit registers. 
Another is a transcendental stream core that can handle 
transcendental functions in addition to the above simple operations.

Moreover, these units are organized hierarchically as follows.
At one level higher from the stream cores, a five-way very long instruction word
unit called a thread processor (TP), that consists of four simple stream cores
and one transcendental stream core.
Therefore, one RV770 processor has 160 TPs.
The TP can execute either at most five single-precision/integer operations, 
four simple single-precision/integer operations with one transcendental operation, 
or double-precision operations by combinations of the four stream cores.
Moreover, a unit called a SIMD engine consists of 16 TPs.
Each SIMD engine has a memory region called a local data store
that can be used to explicitly exchange data between TPs.

\begin{figure}
\centering
\includegraphics[width=13cm,angle=-90]{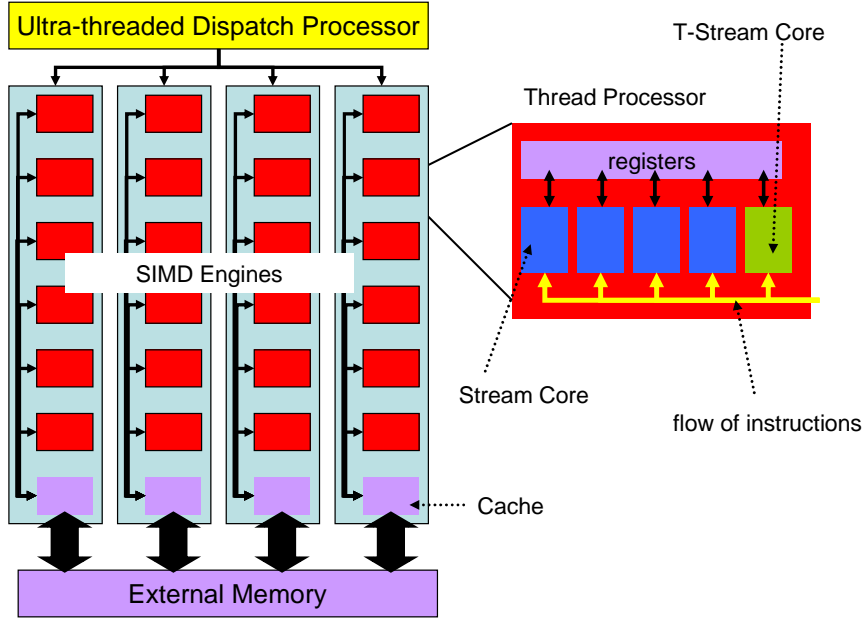}
\caption{A block diagram of RV770 GPU. Note a number of components is not exactly same
as the actual numbers}
\label{RV770}
\end{figure}

At the top level RV770, there are 10 SIMD engines, 
a controller unit called an ultra-threaded dispatch processor, 
and other units such as units for graphic processing, 
memory controllers and DMA engines.
An external memory attached to the RV770 in the present work
is 1 GB GDDR5 memory with a bus width of 256 bit.
It has a data clock rate at 3600 MHz and offers us a bandwidth of 115.2 GB sec$^{-1}$.
In addition to this large memory bandwidth, 
each SIMD engine on RV770 has two-level cache memory.
Figure \ref{RV770} shows a block diagram of RV770.

The RV770 processor with memory chips is mounted on an extension board.
The extension board is connected with a host computer through PCI-Express Gen2 x16 bus.
A theoretical communication speed between the host computer and RV770 GPU is
at most 8 GB sec$^{-1}$ (in one-way).
The measured communication speed of our system is $\sim 5 - 6$ GB sec$^{-1}$ for data size larger than 1 MB.

\subsection{CAL for programming RV770 GPU}
After an introduction of unified shader on GPUs around 2000, 
it became possible to write programs on GPUs by using the shader languages such
as HLSL, GLSL, and Cg.
However those languages were not designed for general computing on GPU.
Even though, an early attempt to implement the force-calculation has been reported \cite{Elsen_2007}.
In 2006, Nvidia inc. provided CUDA (Compute Unified Device Architecture),
which is an integrated development environment of the C/C++ language for GPU programming.
There are many implementations that use CUDA to solve gravitational many-body problem rapidly 
(e.g., \cite{Nyland_2007, Hamada_2007, Belleman_2008}).

In this paper, we used CAL (Compute Abstraction Layer), which is a software development environment
to program GPUs from AMD/ATi Inc.
In CAL, the following API is given.

\begin{itemize}
\item Management of GPU board.
\item Management of resource memory.
\item Making of code for the GPU.
\item Loading and execution of a kernel program.
\end{itemize}

Our program to solve the gravitational many-body problem by using RV770 GPU is composed of two codes.
One is executed with CPU, and the other is executed on the GPU.
The code executed on the GPU is called a kernel. Therefore, the GPU calculates acceleration and
the CPU manages the GPU board with the above API and does numerical integration (updating position and velocity).

\section{Method}
We can analytically solve the two-body problem such as planet movement between the sun and a planet.
However, there are no analytic solutions for problems concerning more than three bodies/particles.
For such problems, only way is to numerically solve $3N$ Newton equations where $N$ is a number of particles.
In the present work, we adopt the leap-frog scheme to integrate the Newton equations.
From time $t_i$ to $t_{i+1}$, it is given by
\begin{eqnarray}
\mbox{\boldmath{$v$}}_{i+1/2}&=&\mbox{\boldmath{$v$}}_i + \frac{1}{2} \mbox{\boldmath{$a$}}_i dt \label{1} \\
\mbox{\boldmath{$x$}}_{i+i}  &=&\mbox{\boldmath{$x$}}_i + \mbox{\boldmath{$v$}}_i dt
                                  + \frac{1}{2} \mbox{\boldmath{$a$}}_i dt^2 \label{2} \\
\mbox{\boldmath{$v$}}_{i+1 } &=&\mbox{\boldmath{$v$}}_{i+1/2} + \frac{1}{2} \mbox{\boldmath{$a$}}_{i+1} dt \label{3}
\end{eqnarray}
where $\mbox{\boldmath{$v$}}$ is velocity of a particle and $dt = t_{i+1} - t_i$ is time step for integration.

The following is an algorithm to solve the general gravitational many-body problem
with (\ref{gravity}), (\ref{1}), (\ref{2}), and (\ref{3}).
\begin{enumerate}
\item Given the initial positions $\mbox{\boldmath{$x$}}$ and velocity $\mbox{\boldmath{$v$}}$
for $N$ particles.
\item Apply (\ref{gravity}) to all particles to compute the acceleration on each particle.
\item Update the velocity of particles with (\ref{1}).
\item Update the position of particles with (\ref{2}).
\item Apply (\ref{gravity}) to all particles then obtain $\mbox{\boldmath{$a$}}_{i+1}$
\item Update the velocity of particles with (\ref{3}).
\item Repeat step 3 to step 6.
\end{enumerate}
In the leap-frog scheme, we need the force-calculation for $N$ particles per one integration timestep.
With reasonably large $N > 1,000$, a most of time consuming part is the exact force-calculation (step 6 above).

\section{Implementation on RV770 GPU}
In this section, we present how do we implement and optimize the exact force-calculation
on RV770 GPU. 
Although RV770 supports double precision operations, 
we use single precision operations throughout the present paper. 

\begin{figure}[tb]
\begin{verbatim}
for i = 0 to N-1
  acc[i] = 0
  for j = 0 to N-1
    acc[i] += f(x[i], x[j])
\end{verbatim}
\caption{A pseudo code for computing the exact force-calculation: a simple two-nested loop}
\label{no-u}
\end{figure}

\begin{figure}[tb]
\begin{verbatim}
for i = 0 to N-1 each 4
  acc[i] = acc[i+1] = acc[i+2] = acc[i+3] = 0
  for j = 0 to N-1
    acc[i]   += f(x[i],   x[j])
    acc[i+1] += f(x[i+1], x[j])
    acc[i+2] += f(x[i+2], x[j])
    acc[i+3] += f(x[i+3], x[j])
\end{verbatim}
\caption{Unroll i-loop in 4 ways}
\label{sim-u}
\end{figure}

\begin{figure}[tb]
\begin{verbatim}
for i = 0 to N-1 each 4
  acc[i] = acc[i+1] = acc[i+2] = acc[i+3] = 0
  for j = 0 to N-1 each 4
    for k = 0 to 3
      acc[i  ] += f(x[i  ], x[j+k])
      acc[i+1] += f(x[i+1], x[j+k])
      acc[i+2] += f(x[i+2], x[j+k])
      acc[i+3] += f(x[i+3], x[j+k])
\end{verbatim}
\caption{Unroll both i-loop and j-loop in 4 ways}
\label{double-u}
\end{figure}

In Figures \ref{no-u}, \ref{sim-u},  \ref{double-u}, 
we present three pseudo codes for computing the exact force-calculation Eq.(\ref{gravity}).
In all cases, \verb+acc[]+ represents gravitational acceleration $\mbox{\boldmath{$a$}}_{i}$
(note we omit the lowest dimension (0,1,2) in \verb+acc[]+ and \verb+x[]+
for brevity but all our code in the present work is three-dimensional).
\verb+f()+ stands for the function $\mbox{\boldmath{$f$}}(\mbox{\boldmath{$x$}}_i, \mbox{\boldmath{$x$}}_j, m_j)$
in Eq. (\ref{gravity}).
Figure \ref{no-u} is a most simple implementation with two-nested loop.
If we simply implement the code like Figure \ref{no-u} on a single-core of a general purpose CPU, 
the calculation of each particle is executed serially.
It is obvious that the loop processing is done $N^2$ times to calculate acceleration of all the particles.
We adopt this code for our CPU implementation and call it {\it only-cpu} from now on.

It is also a simple matter to implement the code like Figure \ref{no-u} on RV770 GPU
but on GPU, calculations of a group of particles are executed in parallel.
With the current our system, logically the calculation for 160 particles are executed in parallel
(note in CAL, details of executions inside GPU is not visible to us so that it is not clear in fact).
We call this implementation {\it gpu-no-unrolling}.
With this code, the two-level cache system on RV770 will be effectively utilized since
 two adjutant processors read \verb+x[j]+ at almost same time.
Namely, once \verb+x[j]+ is stored on the cache by a load instruction issued by a processor,
subsequent load requests to \verb+x[j]+ will be processed very quickly without long latency.

\subsection{Optimization}
In \cite{Elsen_2007}, they have reported their implementation of 
the exact force-calculation of gravity and other forces on an older GPU from AMD/ATi.
A main insight they have obtained was that 
a loop-unrolling technique greatly enhanced the performance of their code.
Loop-unrolling is a technique that reduces the time of loop processing by reducing the loop frequency,
branch instructions and conditional instructions.
We have followed their approach and tried two different ways of the loop-unrolling
shown in Figure \ref{sim-u} and \ref{double-u}.

A code shown in Figure \ref{sim-u} is effective to reduce loop frequency. 
Specifically, in this code, we unroll the outer-loop by four stages.
This enable us not only to reduce the outer loop frequency from $N$ to $N/4$
but also to re-use \verb+x[j]+ repeatedly.
The data re-use is a key optimization in computing on GPU
due to relatively long latency between the GPU chip and the external memory.
We call this implementation {\it simple-unrolling}.
Another key for the optimization is to enhance compute density as much as possible.
For instance, we read one \verb+x[j]+ and compute the function \verb+f()+ four times
in this case.
While with the code in Figure \ref{no-u}, we read one \verb+x[j]+ and compute the function \verb+f()+ one time.
The former case is higher in terms of the compute density.

A code shown in Figure \ref{double-u} is another way of the loop-unrolling.
In this code, we unroll both the inner- and outer-loops by four stages.
From now on, this is called {\it double-unrolling}. 
In this code, we compute 16 force-calculations in the inner-loop
so that the loop frequency is $N^2/16$.
The compute density of this code is further higher than that of the code Figure \ref{sim-u}, 
namely, we read four \verb+x[j]+ and compute the function \verb+f()+ sixteen times.

\section{Experiments}
\subsection{Setup}
To measure a performance of our four implementations (one on CPU and three on RV770 GPU),
we measured the elapsed time for each implementation.
Precisely, we created the following four programs.

\begin{itemize}
\item cpu-only
\item gpu-no-unrolling
\item simple-unrolling
\item double-unrolling
\end{itemize}

We have conducted the following two tests.
\begin{enumerate}
\item Measured the elapsed time with four programs by one step
\item Solved the gravitational many-body problem by 100 timesteps and measured the elapsed time time.
\end{enumerate}

The input data for the tests are as follows.
\begin{itemize}
\item Three figure-eight galaxies
\begin{itemize}
\item $N =$7500, timestep $dt=0.015625$ and $\epsilon = 0.1$.
This input data have initial position and velocity for each particle as the orbit of three galaxies draw figure eight. 
\end{itemize}

\item Four figure-eight galaxies
\begin{itemize}
\item $N = 10000$, timestep $dt=0.015625$ and $\epsilon = 0.1$. 
This input data have initial position and velocity for each particle as the orbit of four galaxies draw figure eight.
\end{itemize}

\item A cart-wheel galaxy
\begin{itemize}
\item $N = 18000$, timestep $dt=0.0078125$ and $\epsilon = 0.01$.
This input data have initial position and velocity for each particle
as it becomes a wheel galaxy by colliding two galaxies.
\end{itemize}

\item Random 100000-body
\begin{itemize}
\item $N = 100000$, timestep $dt=0.01$ and $\epsilon = 0.01$. 
This input data have random initial position and velocity.
\end{itemize}
\end{itemize}

We have used a configuration shown in Table \ref{win} for our experiments shown here.

\begin{table}
\renewcommand{\arraystretch}{1.3}
\caption{Our windows configuration}
\label{win}
\begin{center}
\begin{tabular}{c|c}
\hline
& Part  \\
\hline
CPU & Core2 E8400  \\
MB  & Dell Precision T3400  \\
Memory & DDR2 800 1GB x2 \\
GPU & Radeon HD4580 512MB (core clock 625 MHz, memory 512 MB) \\
OS & Windows Vista SP1 (64 bit) \\
Compiler & Visual C++ 2008 Express Edition \\
API & OpenGL (for visualization) \\
CAL ver. & 1.2 \\
Catalyst ver. & 8.9 \\ 
\hline
\end{tabular}
\end{center}
\end{table}

\subsection{Results}
\begin{figure}
\begin{center}
\includegraphics[width=100mm]{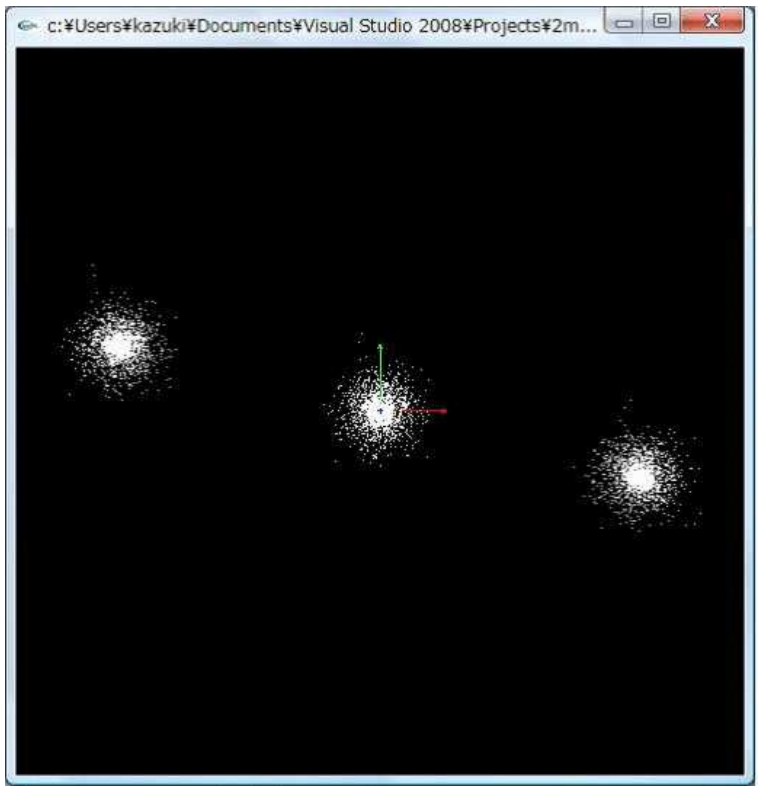}
\end{center}
\caption{Initial snapshot of 7500-body galaxy}
\label{fig1}
\end{figure}

\begin{figure}
\begin{center}
\includegraphics[width=100mm]{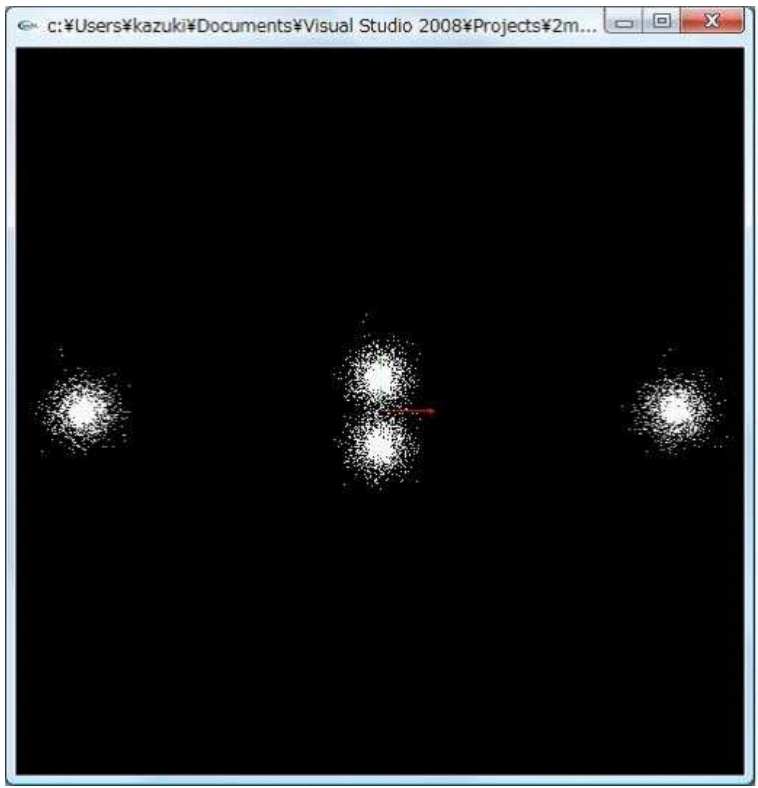}
\end{center}
\caption{Initial snapshot of 10000-body galaxy}
\label{fig2}
\end{figure}

\begin{figure}
\begin{center}
\includegraphics[width=100mm]{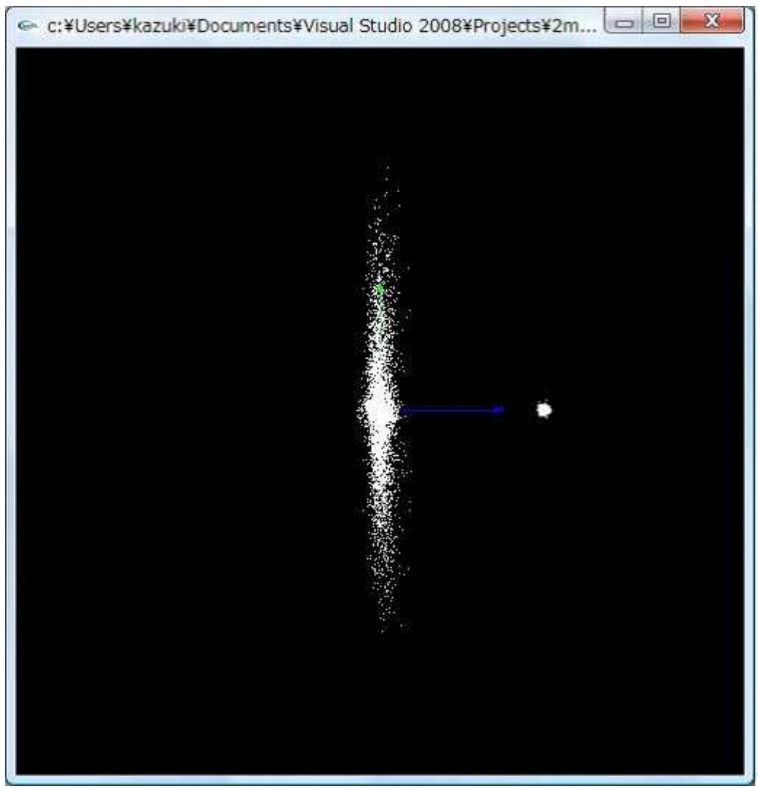}
\end{center}
\caption{Initial snapshot of 18000-body galaxy (side view)}
\label{fig3a}
\end{figure}

\begin{figure}
\begin{center}
\includegraphics[width=100mm]{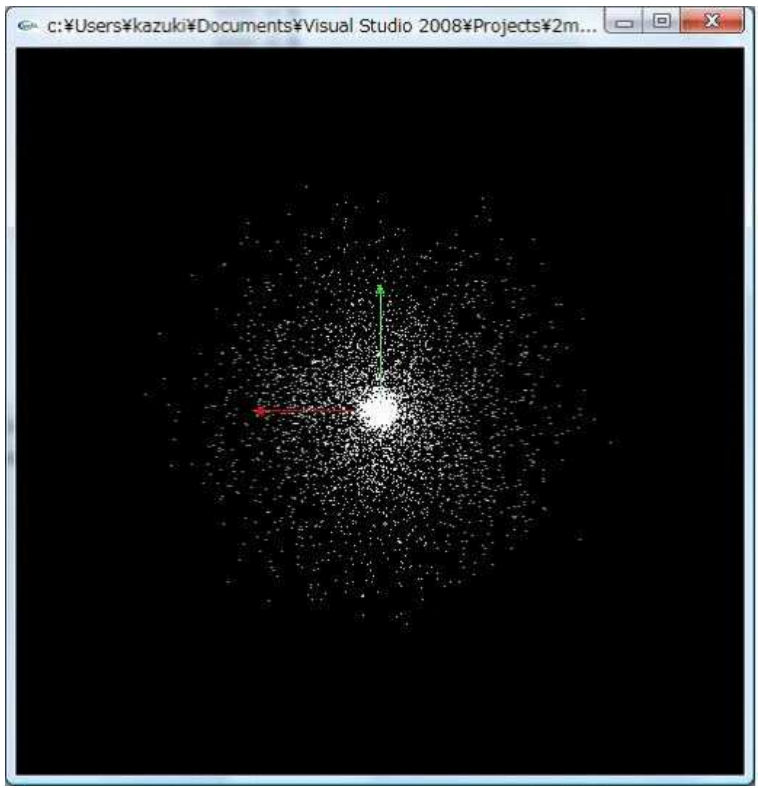}
\end{center}
\caption{Initial snapshot of 18000-body galaxy (front view)}
\label{fig3b}
\end{figure}

\begin{figure}
\begin{center}
\includegraphics[width=100mm]{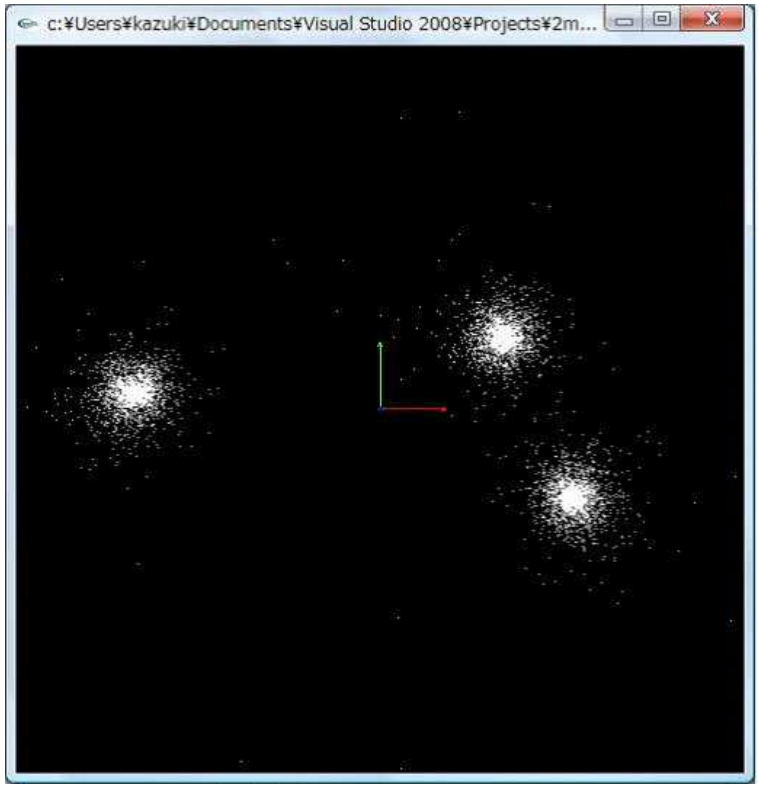}
\end{center}
\caption{A snapshot of 7500-body galaxy}
\label{fig1-e}
\end{figure}

\begin{figure}
\begin{center}
\includegraphics[width=100mm]{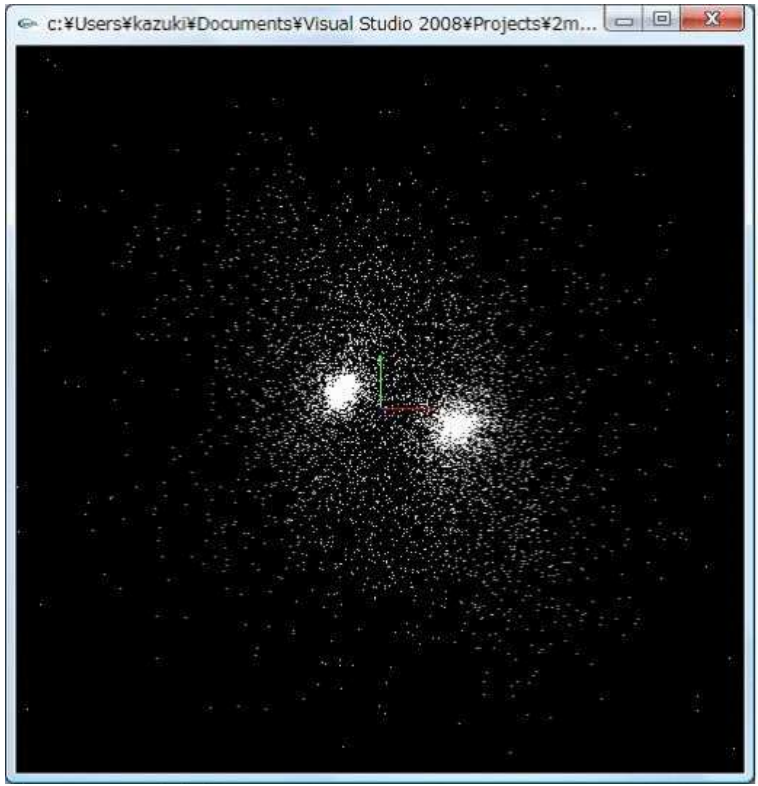}
\end{center}
\caption{A snapshot of 10000-body galaxy}
\label{fig2-e}
\end{figure}

\begin{figure}
\begin{center}
\includegraphics[width=100mm]{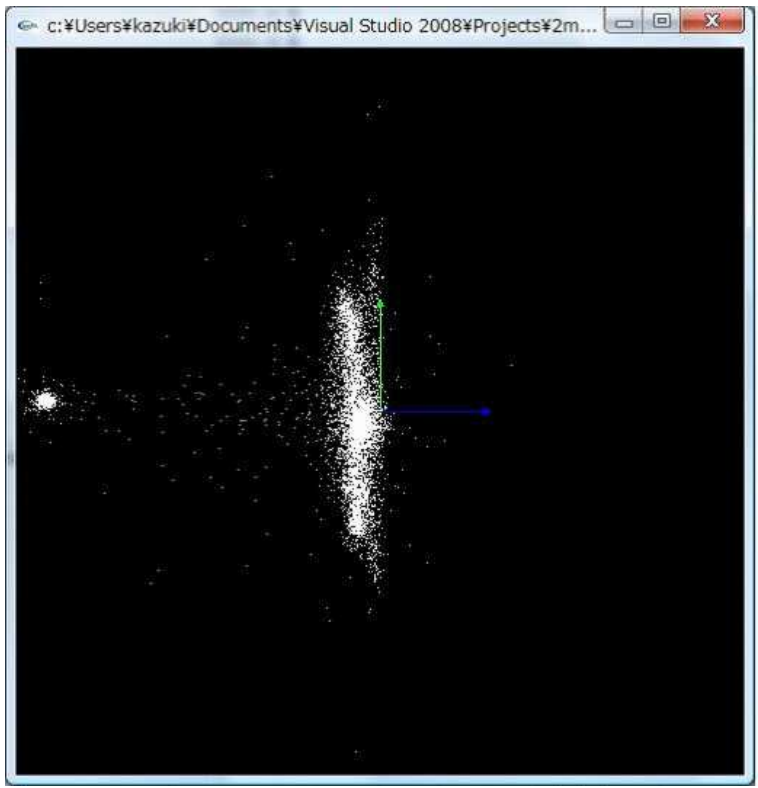}
\end{center}
\caption{A snapshot of 18000-body galaxy (side view)}
\label{fig3a-e}
\end{figure}

\begin{figure}
\begin{center}
\includegraphics[width=100mm]{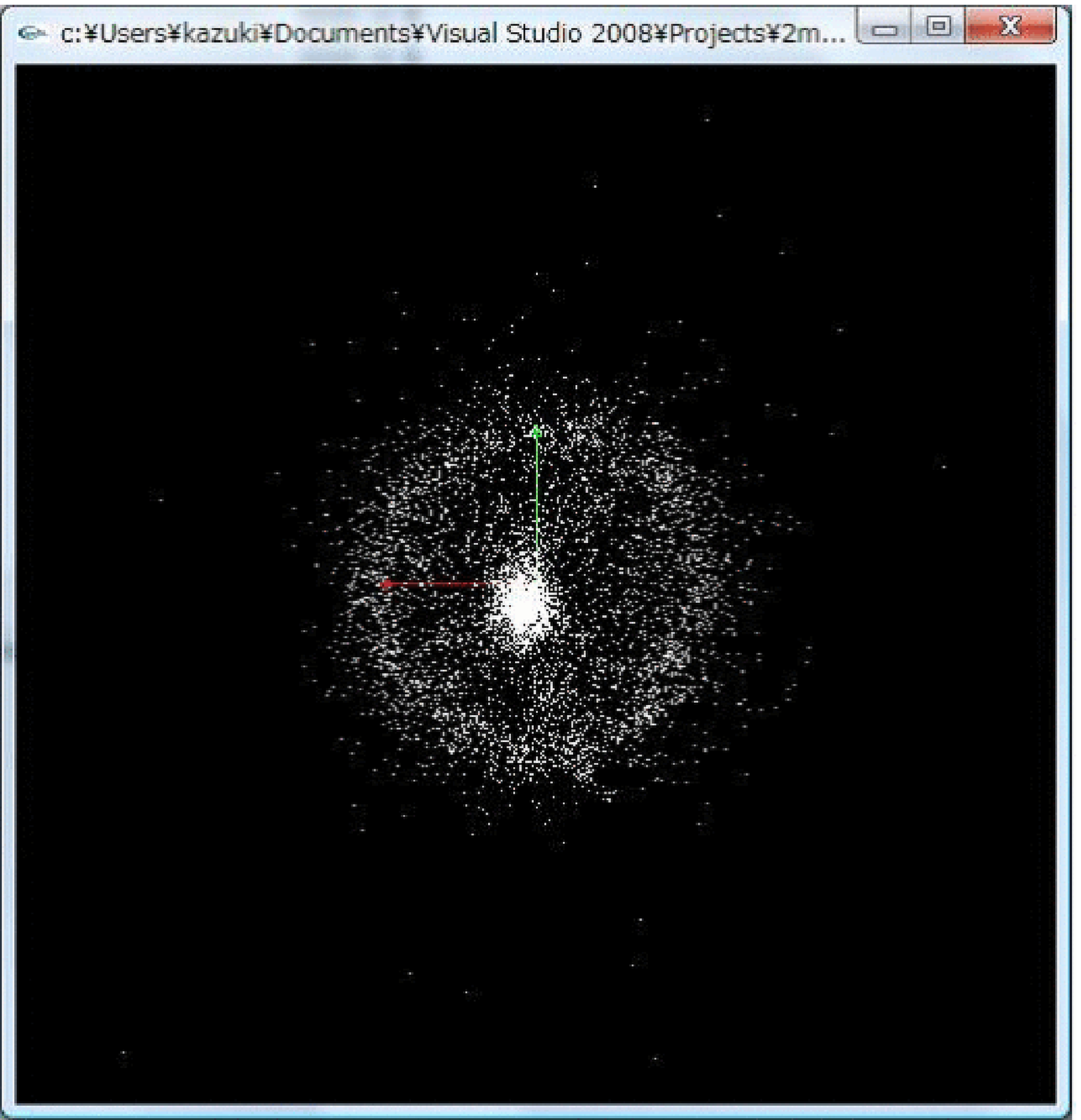}
\end{center}
\caption{A snapshot of 18000-body galaxy (front view)}
\label{fig3b-e}
\end{figure}

Figure \ref{fig1}, \ref{fig2}, \ref{fig3a} and \ref{fig3b}
show initial snapshots for 7500-body, 10000-body, 18000-body (side view) and
18000-body (front view) galaxies, respectively,
In the case of 18000-body galaxy, a small blob in Figure \ref{fig3a} represents
an intruder to the disk galaxy at the center.
It moves to left in this Figure.
Figure \ref{fig1-e} and \ref{fig2-e} shows snapshots after 100 timesteps
for 7500-body and 10000-body galaxies, respectively.
Figure \ref{fig3a-e} and \ref{fig3b-e} shows snapshots after 100 timesteps
for 18000-body (side view) and 18000-body (front view) galaxies, respectively,

Table \ref{ex1} shows the elapsed time for Experiment 1.
It is clear that the elapsed time of the three programs using the GPU is
much shorter than that of the cpu-only program.
The gpu-no-unrolling is faster than the cpu-only by a factor of 400 to 800.
Furthermore, the simple-unrolling is faster than the cpu-only by a factor of 500 to 1200.
Finally, the double-unrolling is much faster than the cpu-only by a factor of 600 to 1900.

Table \ref{ex2} shows the elapsed time for Experiment 2.
We see that the loop-unrolling technique on RV770 GPU are very effective.
Depending on $N$, the double-unrolling is faster than the simple-unrolling
by a factor of 1.08, 1.35, 1.28, and 1.45 for $N = 7500, 10000, 18000, 100000$, respectively.

\begin{table}
\renewcommand{\arraystretch}{1.3}
\begin{center}
\begin{tabular}{l|r|r|r|r}
\hline
program / particle& 7500 & 10000 & 18000 & 100000 \\
\hline\hline
cpu-only & 5.6 & 9.7 & 32 & 971 \\
gpu-no-unrolling & 0.012 & 0.023 & 0.052 & 1.19 \\
simple-unrolling & 0.010 & 0.016 & 0.035 & 0.78 \\
double-unrolling & 0.009 & 0.011 & 0.029 & 0.50 \\
\hline
\end{tabular}
\end{center}
\caption{The elapsed time for one step (in second)}
\label{ex1}
\end{table}

\begin{table}
\renewcommand{\arraystretch}{1.3}
\begin{center}
\begin{tabular}{l|r|r|r|r}
\hline
program / particle & 7500 & 10000 & 18000 & 100000 \\
\hline\hline
gpu-no-unrolling & 0.95 & 1.19 & 4.77 & 119 \\
simple-unrolling & 0.74 & 1.24 & 3.10 & 79.8 \\
dividing-input & 0.68 & 0.92 & 2.43 & 55.0 \\
\hline
\end{tabular}
\end{center}
\caption{The elapsed time for integrating each system for 100 timesteps (in second)}
\label{ex2}
\end{table}

\begin{figure}[!h]
\includegraphics[width=150mm]{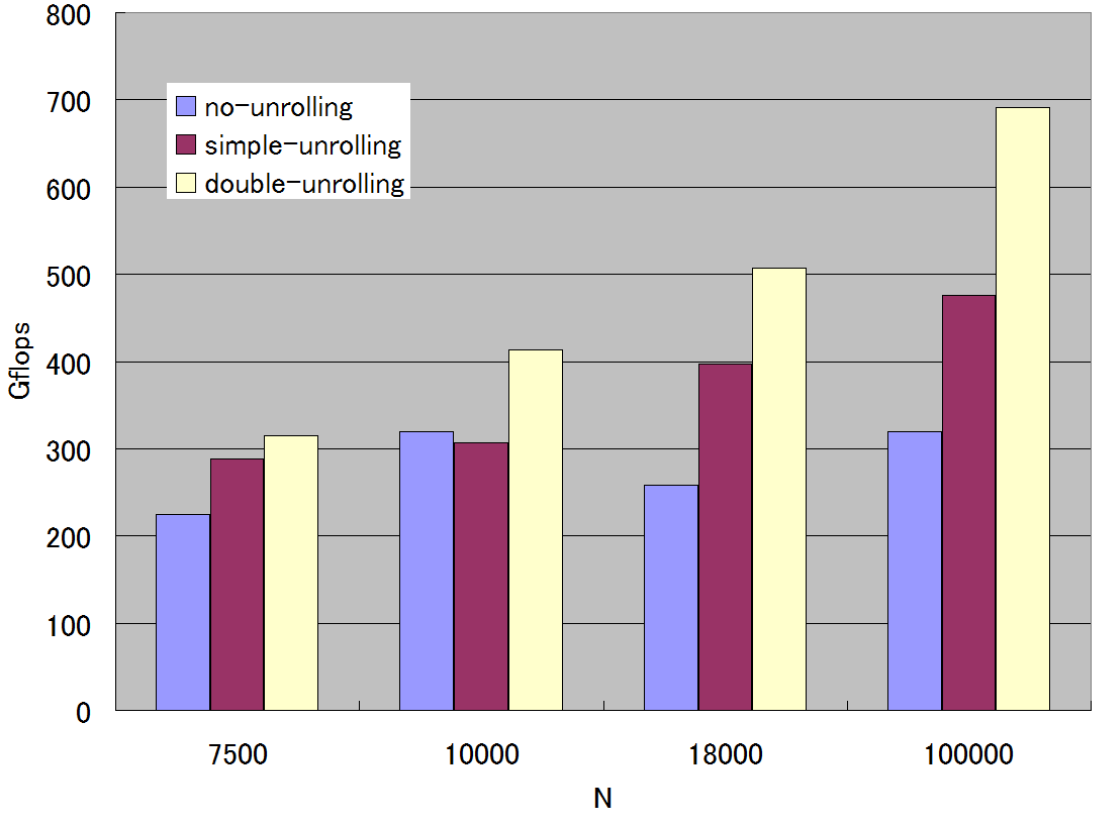}
\caption{Performance of three codes for RV770 GPU obtained with Experiment 2 as a function of N}
\label{p}
\end{figure}

Finally, Figure \ref{p} shows the computing performance of our three GPU implementations.
In this figure, we adopted 38 floating operations per one interaction
to compute the performance of our codes.
From this figure, the largest calculating speed to solve the gravitational many-body problem
on the GPU is 319 Gflops with the gpu-no-unrolling program, 
that is 476 Gflops with the simple-unrolling program
and that is 690 Gflops with the double-unrolling program.

\section{Discussion}
We have used GPU ShaderAnalyzer, which is provided by AMD/ATi Inc.
to see the usage of the number of registers and ALU instructions on a GPU kernel, 
to analyze our implementations.
It turned out that the gpu-unrolling used only 6 registers and 11 ALU instructions.
Because of higher compute density, the simple-unrolling program used 10 registers and 25 ALU instructions
and the double-unrolling program used 34 registers and 72 ALU instructions.
It is expected that utilizing more many registers, a number of data load is smaller
so that the performance of the double-unrolling is fastest among three implementations.

\begin{figure}
\centering
\includegraphics[width=12cm,angle=-90]{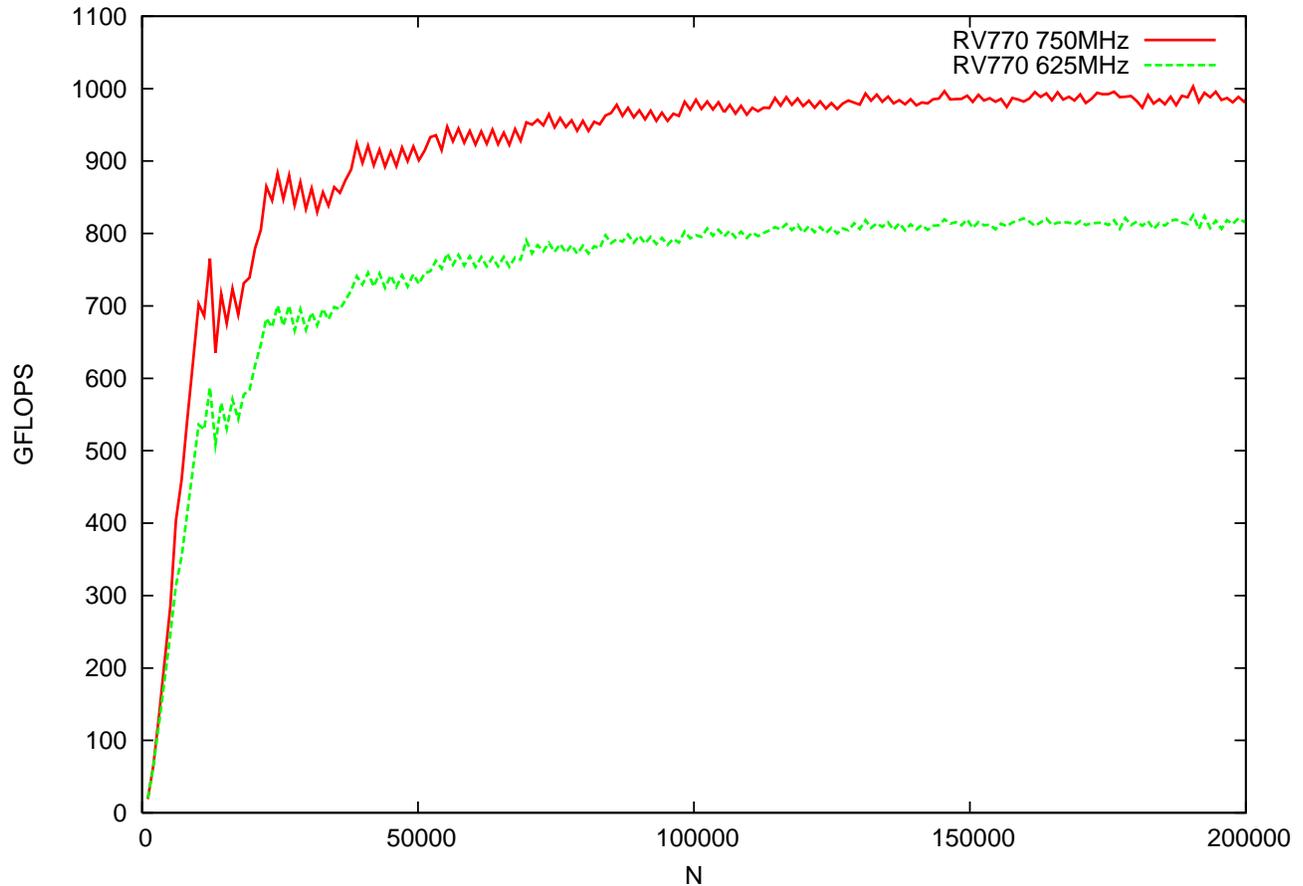}
\caption{A performance of the optimized code on two types of RV770 GPU}
\label{bare}
\end{figure}

In the conducted experiments, we have tested only restricted variation of $N$.
In addition, there are many possible detailed optimizations to further enhance the performance of
the double-unrolling program.
After extensive optimizations, 
the highest performance obtained so far is $\sim 1$ Tflops as shown in Figure \ref{bare}.
In Figure \ref{bare}, we plot a computing speed of our optimized code for computing Eq.(\ref{gravity})
as a function of $N$. 
For this particular test, we have used a different configuration as shown in Table \ref{conf}.
We have tested two configurations as one RV770 GPU running at 625 (HD4850) and
750 MHz (HD4870), respectively.
So far, we have obtained a maximum performance of $\sim 990$ GFLOPS with $N \sim 200,000$. 
With $N = 226,304$, our optimized brute force method took roughly 2 seconds on one RV770 running at 750MHz.
As far as we know, the performance we obtained is fastest ever with one GPU chip in April 2009.

\begin{table}
\renewcommand{\arraystretch}{1.3}
\caption{Our optimized configuration}
\label{conf}
\begin{center}
\begin{tabular}{c|c}
\hline
& Part  \\
\hline
CPU & Core2 E8400  \\
MB  & ASUS P5E WS    \\
Memory & DDR2 800 1GB x4 \\
GPU & Radeon HD4780 1GB/HD4580 512MB \\
OS & Ubuntu 8.04 LTS (x86-64) \\
CAL ver. & 1.4   \\
Catalyst ver. & 9.3  \\ 
\hline
\end{tabular}
\end{center}
\end{table}

A computing performance of other hardware-accelerators is as follows.
The performance of GRAPE-5 \cite{Kawai_2000} is about 40 Gflops
and that of GRAPE-7 model 800, which is the newest variant of GRAPE system, is $\sim 800$ Gflops.
An already reported computing performance of Nvidia GeForce 8800 GTX \cite{Nyland_2007}
is $\sim$ 370 Gflops. They have used CUDA to program their GPUs.
RV770 GPU that we have used in the present paper (Radeon HD4850/4870)
is a comparatively new in 2009.
That is a raw higher performance of HD4870 unable us
to obtain the fastest computation of the exact force-calculation.
On the other hand, internal architecture of G80 GPU from Nvidia is quite different from RV770 GPU.
We speculate that the cache system on RV770 GPU is very effective to
the exact force-calculation of gravity.
We think lack of a cache system on G80 is critical difference over RV770 GPU.

\section{Conclusion}
In this paper, we present our implementation of the exact force-calculation of gravity for on RV770 GPU.
We have obtained significantly improved performance by efficiently utilizing the GPU.
Moreover, the calculation performance on the GPU is sensitively depending on how we do the loop-unrolling.  
In the present work, we have implemented only two types of a loop-unrolling technique.
Apparently, it is expected that a number of stages to unroll affects the computing performance.
This will be a possible future work.
In addition, investigation of more sophisticated algorithms with better computational complexity
on RV770 GPU will be another interesting future work.

\end{document}